\documentclass[aps,pra,twocolumn,showkeys,showpacs,a4paper,nobibnotes,superscriptaddress]{revtex4-2}
\usepackage[T1]{fontenc}
\usepackage{mathptmx}
\usepackage{graphicx}
\usepackage{amsmath}
\usepackage[colorlinks,citecolor=blue,linkcolor=blue,urlcolor=blue,bookmarks=true]{hyperref}
\usepackage{soul}

\graphicspath{{figures/pdf/}{figures}}

\begin{document}

\title{Association between projectile and target excitation in slow \mbox{$\text{Ar}^{q+}\text{--\,CO}_2$} collisions}

\author{Akash Srivastav}
\email[]{akash.srivastav@students.iiserpune.ac.in}
\affiliation{Indian Institute of Science Education and Research Pune, Homi Bhabha Road, Pune~411008, India}
\author{Sumit Srivastav}
\affiliation{Indian Institute of Science Education and Research Pune, Homi Bhabha Road, Pune~411008, India}
\affiliation{\textup{Present address:} Normandie Univ., ENSICAEN, UNICAEN, CEA, CNRS, CIMAP, 14000 Caen, France}
\author{Vishnu P}
\author{Bhas Bapat}
\email[]{bhas.bapat@iiserpune.ac.in}
\affiliation{Indian Institute of Science Education and Research Pune, Homi Bhabha Road, Pune~411008, India}

\date{\the\day.\the\month.\the\year}

\begin{abstract}

We investigate ionic fragmentation of CO$_2^{n+}$~\mbox{($2\le n\le 4$)} produced in collisions with Ar$^{q+}$~\mbox{($4\le q\le 16$)} projectiles at a collision velocity of $\approx$~0.3~a.u.  For most projectile and fragmentation channel combinations, the shape of the kinetic energy release distribution (KERD) differs with the electron capture mediated charge change (\mbox{$\Delta q$}) in the scattered projectile: KERD for \mbox{$\Delta q = 2$} is broader at high KER than for \mbox{$\Delta q =1$}. The difference generally diminishes with increasing projectile charge.  Two deviations in this general trend are seen in the fragmentation of CO$_2^{3+}$, one for Ar$^{4+}$ impact in the high KER region and the other for Ar$^{6+}$ impact in the low KER region. The calculated reaction windows for multielectron capture within the framework of the extended classical over-the-barrier model (ECOBM) indicate that while ionization of the target occurs via multielectron capture, the scattered projectile may subsequently undergo multi-fold autoionization. Interpreting projectile autoionization to be a consequence of capture into highly excited states and high fragment KER to be a consequence of excitation of the ionized target to high-lying states, we find a strong dependence between the target and scattered projectile excitations. 

\end{abstract}


\maketitle

\section{INTRODUCTION}
\label{sec:intro}

Electron capture is the dominant ionization mechanism for slow collisions (\mbox{$v<1$}~a.u.) \cite{Vancura_1994,Wu_1995,Wells_2005,Mawhorter_2007,Luna_2016,Wu_2025}. In this regime, it is therefore reasonable to assume that $r$-fold ionization of the target is predominantly initiated by the capture of $r$ electrons by the projectile. The captured electrons typically populate excited states of the scattered projectile, which may then relax via photon emission, referred to as radiative stabilization, or via autoionization, thus losing some of the captured electrons.  In the latter case a projectile of initial charge $q$ that captures $r$ electrons and undergoes $s$-fold autoionization will have a final charge $q-r+s$.  Irrespective of the method of relaxation, capture-mediated collisions serve as effective probes of multiply-excited atoms and molecules \cite{Stolterfoht_1986,M_Barat_1992,Stolterfoht_1994,Fremont_1996,Sobocinski_2001}.

We will denote capture processes by the label $r$C$s$A$_\text{P}$, with $r$ the number of electrons captured and $s$ the number lost by projectile autoionization. Radiative stabilization is pure capture and will be denoted by $r$C. In most cases under slow collisions, $r$ is also the degree of ionization of the target. It may also happen that some ($r$) electrons are captured and some ($i$) electrons are released to the continuum. This process is sometimes called transfer ionization.  We will denote this process by $r$C$i$I$_\text{T}$; the degree of target ionization in this case will be \mbox{$r+i$}.  Furthermore, capture may be accompanied by excitation of other target electrons \cite{Hoekstra_1993,Hasan_1999,Song_2020}. If this leads to $s$-fold target autoionization, the target charge state will change to \mbox{$r+s$}. We denote this process by $r$C$s$A$_\text{T}$. Additionally, direct ionization (DI) can occur, wherein one or more target electrons are ejected directly into the continuum without any change in the projectile’s charge state.

Within the framework of the extended classical over-the-barrier model (ECOBM) \cite{Niehaus_1986} for ion--atom collisions, distant collisions preferentially lead to capture of the outermost target electrons, resulting in comparatively low excitation of the target ion while populating high-lying multiply excited states of the scattered projectile. In contrast, close collisions enable electron transfer from deeper shells, imparting higher excitation to the target and leading to the population of more tightly bound electronic states in the scattered projectile.  Highly excited states of the scattered projectile may decay via autoionization, provided the total energy of the configuration exceeds the corresponding autoionization threshold.

If the target is a molecule, the excitation arising from the collision may render the molecular ion unstable and it may dissociate into charged and/or neutral fragments.  The net effect of the capture and excitation processes gets reflected in the distribution of the fragmentation products and their kinematics.   The kinetic energy release (KER), defined as the sum of the kinetic energies of the fragments, and its distribution, KERD, serve as a sensitive probe of electronic state populations and energy deposition in the molecular ion.

The role of capture and associated processes in ionization and fragmentation of molecular ions has been investigated extensively across slow (\mbox{$v < 1$}~a.u.) \cite{Folkerts_1996,Folkerts_1997,Ehrich_2003,Tezuka_2018,Azad_2022,Azad_2023,Kumar_2024_a,Azad_2025,Akash_2025}, intermediate (\mbox{$v\approx 1$}~a.u.) \cite{Wang_2021,Azad_2023,Gao_2023,Kumar_2024_b,Azad_2025}, and fast (\mbox{$v>1$}~a.u.) collision regimes \cite{Mizuno_2010_a,Mizuno_2010_b,Lamichhane_2017}. In the slow collision regime several studies have reported a correlation between the degree of excitation in the scattered projectile and that in the ionized target \cite{Folkerts_1996,Folkerts_1997,Ehrich_2003,Tezuka_2018,Azad_2022,Azad_2023,Kumar_2024_a,Azad_2025}.  In 0.35~a.u. Ar$^{8+}$–CO$_2$/OCS collisions, Tezuka \emph{et al.} \cite{Tezuka_2018} observed enhancement of the intensity in the high KER region for capture followed by projectile single autoionization compared to double autoionization. Similar trends were reported for Ar$^{6+}$ and Ar$^{8+}$ projectiles with various targets \cite{Azad_2022,Azad_2023,Kumar_2024_a}. Although the influence of capture-associated processes has been explored for various degrees of target ionization, the projectile charge was either fixed \cite{Tezuka_2018,Azad_2023} or varied over a limited range \cite{Ehrich_2003}. Conversely, studies with varying projectile charge focused on a fixed degree of target ionization \cite{Kumar_2024_a}. A comprehensive investigation spanning a broad range of projectile charge states and target ionization degrees is still lacking.

We present here a study of the fragmentation dynamics of CO$_2^{n+}$~\mbox{($2 \le n \le 4$)} produced in collisions with Ar$^{q+}$~\mbox{($4 \le q \le 16$)} projectiles at collision velocities of 0.27~a.u. for Ar$^{4+}$ and 0.31~a.u. for Ar$^{q+}$~\mbox{$(q \ge 6)$}.  Measurements are performed with post-collision charge-state analysis of the scattered projectile, for the cases where the charge of projectile reduces by \mbox{$\Delta q =1,2$}.  KERDs are reported for ionic fragmentation channels for target charge states \mbox{2--4}.  KERDs corresponding to \mbox{$\Delta q =1,2$} show differences with certain trends, and the differences diminish with increasing projectile charge. The results are qualitatively interpreted using reaction windows derived within the ECOBM framework.

\section{EXPERIMENTAL DETAILS}
\label{sec:exp}

Details of the experimental setup have been described elsewhere \cite{Bapat_2020,Vandana_2006,Sumit_2022_CDA}; only a brief overview of the experimental scheme is provided here. The projectile ion beam was delivered by the electron beam ion source (EBIS) facility at IISER Pune \cite{Bapat_2020}. The ion beam intersected an effusive CO$_2$ gas jet at the center of a multihit-capable ion momentum spectrometer (IMS) \cite{Vandana_2006}, operated under Wiley–McLaren space-focusing conditions \cite{Wiley_1955}.

A uniform extraction field of 60~V/cm, oriented perpendicular to both the projectile beam and the gas jet, guided the recoil and fragment ions toward a position-sensitive ion detector. Electrons ejected during the collision were simultaneously guided to a channeltron detector mounted opposite the ion detector. Specific charge-changed projectiles were transported to a channeltron detector positioned at the exit slit of a cylindrical deflector analyzer coupled downstream of the IMS.

The present experiments involved Ar$^{q+}$~\mbox{($4 \le q \le 16$)} projectiles colliding with CO$_2$ molecules at velocities close to 0.3~a.u.: \mbox{$v=0.27$}~a.u.\ for $q=4$ and \mbox{$v=0.31$}~a.u.\ for \mbox{$q \ge 6$}. The chamber pressure during operation was maintained at a few $10^{-7}$~mbar, and the typical ion beam current was $\approx$ 10~pA.

In the present experiments, fragment and recoil ions produced in the collision were detected in coincidence with charge-changed projectiles. Experiments are restricted to \mbox{${\Delta q=1}$} and \mbox{${\Delta q=2}$}.

\section{CAPTURE-ASSOCIATED PROCESSES and THE ECOBM}
\label{sec:reacwin}

For the collision velocity considered in this work (\mbox{$v < 1$}~a.u.), multielectron capture is the dominant process over the entire range of projectile charge states studied; other contributions can be ignored.  This conclusion is consistent with the classical Bohr-Lindhard model \cite{Bohr_1954}, in which ionization of electron(s) to the continuum becomes energetically allowed only for collision velocities exceeding \mbox{$v_{\text{min}} = q^{1/4} I^{1/2}$}, where $q$ is the projectile charge and $I$ is the ionization energy of the target electron (in a.u.).  Experiments have confirmed this velocity threshold for double DI and 1C1I$_\text{T}$ processes \cite{Wu_1995,Wells_2005}.  For the lowest projectile charge (\mbox{$q=4$}) considered in this work, the onset of ionization is expected for \mbox{$v \ge v_{\text{min}} = 1$~a.u.}, which is significantly above the collision velocity employed here. Therefore, we take multielectron capture as the predominant mechanism leading to the formation of CO$_2^{r+}$.

The ECOBM has been extensively employed in the literature to qualitatively elucidate the role of capture-associated processes in the sharing of internal energy between the collision partners \cite{Hoekstra_1993,Folkerts_1996,Folkerts_1997,Tezuka_2018,Azad_2023,Kumar_2024_a,Kumar_2024_b,Azad_2025}. Within this framework, a capture event is represented by a string; for example, the string \mbox{$(j)$ = (1,0,1,1)} denotes that four target electrons, starting from the outermost valence electron, became quasi-molecular (i.e.\ bound in the combined projectile–target nuclear Coulomb field) during the collision. Of these, the first, third, and fourth electrons were captured by the projectile, whereas the second electron was recaptured by the target. The model also incorporates transfer-excitation, whereby a recaptured electron populates excited state of the target, provided that the capture of an inner electron has occurred. Accordingly, the degree of excitation imparted to the target increases as the position of the recaptured electron shifts to the left in the string. For instance, in the string (1,1,1,0) the fourth electron is recaptured into the ground state of the triply ionized target, whereas the string (0,1,1,1) corresponds to the highest degree of excitation of the ionized target.

For a given string \mbox{($j$)}, let $E^{(j)}$ denote the sum of the binding energies of all captured electrons. In the static limit, ECOBM predicts a discrete value \mbox{$E^{(j)} = E_0^{(j)}$}. To account for the finite collision velocity, however, the model assumes that the realized binding energy $E^{(j)}$ follows a Gaussian probability distribution centered at $E_0^{(j)}$:
\begin{align}
	W(E^{(j)}) &= \frac{1}{\Delta E^{(j)}\pi^{1/2}}\text{exp}\left( - [( {E^{(j)}-E_0^{(j)}})/{\Delta E^{(j)}}]^2 \right)
	\label{eq1}
\end{align}
where
\begin{align}
	(\Delta E^{(j)})^2 &= \sum_{t} (\Delta E_t^{(j)})^2 \label{eq2}
\end{align}
and
\begin{align}
	E_0^{(j)} &= \sum_{t} (\epsilon_t^{(j)}) \label{eq3}
\end{align}
where the index $t$ runs over all captured electrons.  The expressions for the energy width $\Delta E_t^{(j)}$ associated with the predicted binding energies $\epsilon_t^{(j)}$ of a captured electron are provided in Ref.~\cite{Niehaus_1986}.

The ionization energies of CO$_2$ required for the present analysis were calculated using the coupled cluster singles and doubles (CCSD) method with Dunning-type correlation-consistent basis sets augmented with diffuse functions (aug-cc-pVTZ), within the GAMESS suite of programs \cite{Schmidt_1993}. The first four ionization energies of CO$_2$ were obtained as 13.78, 23.79, 37.00, and 48.08~eV, respectively. The probability distributions of the total binding energy of the captured electrons, as defined above, are commonly referred to as \emph{reaction windows} in the literature.  Reaction windows for specific strings are calculated and compared with the calculated range of binding energies of the captured electrons in the scattered projectile.  The binding energies are calculated using the flexible atomic code (FAC) \cite{Gu_2008}.

In calculating the reaction windows relevant to the formation of CO$_2^{r+}$, we consider collisions initiated by the capture of $r$ electrons, assuming that \mbox{$r+1$} electrons become quasi-molecular during the interaction. Under this assumption, $r+1$ capture strings are realized. In the following discussion, we examine the ranges of these reaction windows relative to the $s$-fold autoionization threshold ($A_s$) of the scattered projectile. The portion of the reaction windows lying above $A_{r-1}$ contributes to \mbox{$\Delta q=1$} via the $r$C$(r-1)$A$_\text{P}$ process, while that lying above $A_{r-2}$ contributes to \mbox{$\Delta q=2$} via the $r$C$(r-2)$A$_\text{P}$ process. If a substantial fraction of the reaction windows lies above the corresponding thresholds, these processes are strongly favored and predominantly contribute to \mbox{$\Delta q=1,2$}.

\section{OBSERVED KER DISTRIBUTIONS}
\label{sec:kerd}

We report KERDs for two- and three-body ionic fragmentation channels of CO$_2^{r+}$~\mbox{($2 \le r \le 4$)}. The objective of the present analysis is to gauge the correlation between the excitation of the projectile due to different capture-associated processes and the energy deposition in the target leading to multiple ionization and fragmentation.  We also examine how the projectile charge $q$ influences the relative importance of various capture-associated processes for a fixed degree of target ionization.

For consistency and meaningful comparison, KERDs for a given fragmentation channel are presented only for those projectile charge states for which sufficient statistics were obtained for both \mbox{${\Delta q=1}$} and \mbox{${\Delta q=2}$}.

\subsection{Ionic Fragmentation of CO$_2^{2+}$}
\label{sec:frag2}

\begin{figure*}
\centering
\includegraphics[width=0.8\textwidth]{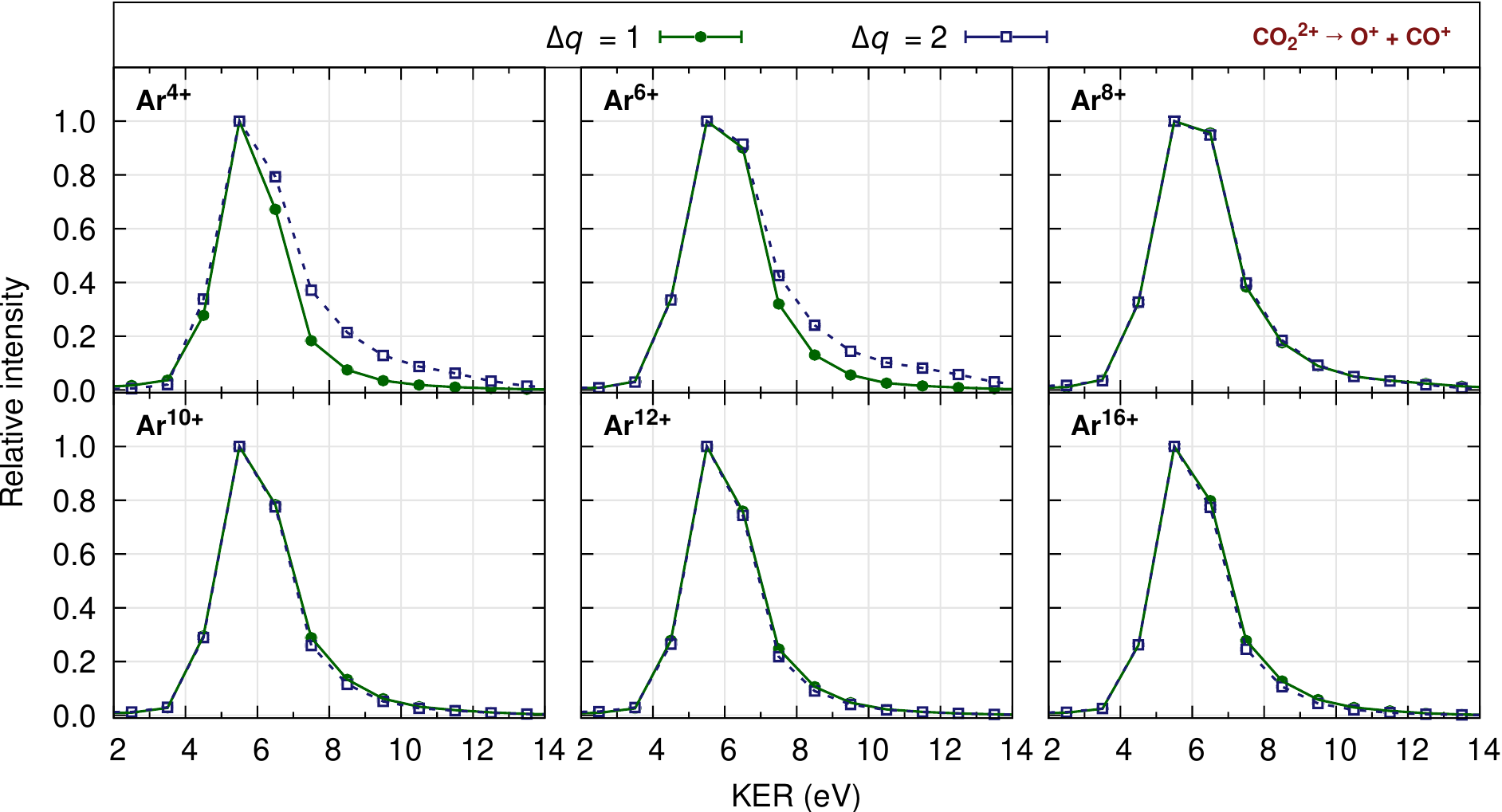}
\caption{KERDs for the \mbox{O$^+$:\,CO$^+$} break-up channel of CO$_2^{2+}$, resulting from the impact of various Ar$^{q+}$ projectiles undergoing a charge change, \mbox{$\Delta q= 1,2$}. Error bars represent statistical uncertainties.}
\label{fig:ker22}
\end{figure*}

CO$_2^{2+}$ undergoes ionic fragmentation only via the two-body channels \mbox{O$^+$:\,CO$^+$} and \mbox{C$^+$:\,O$_2^+$}. Although the isomerization channel \mbox{C$^+$:\,O$_2^+$} is observed as a weak signal for all projectiles for \mbox{$\Delta q =1$}, it is seen for both \mbox{$\Delta q= 1,2$} collisions only for Ar$^{4+}$ impact, and is consequently excluded from our discussion.

KERDs for the \mbox{O$^+$:\,CO$^+$} break-up channel are shown in Fig.~\ref{fig:ker22}. The distributions exhibit a single prominent peak centered around $\sim$~5.5~eV for both \mbox{$\Delta q= 1$} and \mbox{$\Delta q = 2$}. This peak agrees well with earlier studies of the same fragmentation channel induced by intense laser fields \cite{Larimian_2017}, electron impact \cite{Sharma_2007,Bhatt_2012,Wang_2013}, and proton impact \cite{Duley_2023}.

There are differences in the KERDs for \mbox{$\Delta q = 1,2$} in the case of  Ar$^{4+}$ and Ar$^{6+}$ impact. For these two projectiles, the distributions corresponding to \mbox{$\Delta q=2$} exhibit a small enhancement in the high-KER region (beyond $\sim$~6~eV) compared to those corresponding to \mbox{$\Delta q= 1$}, while the two distributions become nearly identical at lower KER values.  As the projectile charge increases, the KERDs for \mbox{$\Delta q= 1,2$} become virtually identical.  These observations are consistent with that of Tezuka \emph{et al.} \cite{Tezuka_2018} for two-body fragmentation of CO$_2^{2+}$ and OCS$^{2+}$ following Ar$^{8+}$ impact.

\subsection{Ionic Fragmentation of CO$_2^{3+}$}
\label{sec:frag3}

We first summarize the observed two-body ionic fragmentation channels, \mbox{O$^+$:\,CO$^{2+}$} and \mbox{O$^{2+}$:\,CO$^+$}. The former is observed for both \mbox{$\Delta q= 1$} and \mbox{$\Delta q= 2$}, but only for \mbox{$q\ge 8$}, while the latter for \mbox{$q\ge 10$}. For both channels, the KERDs corresponding to \mbox{$\Delta q= 1,2$} are, by and large, similar for $q \ge 10$. However, for the \mbox{O$^+$:\,CO$^{2+}$} channel under Ar$^{8+}$ impact, the high-KER region exhibits a slight enhancement in intensity for \mbox{$\Delta q =2$} compared to \mbox{$\Delta q =1$}, while at low KER values the two are nearly identical. This behavior is analogous to that observed for the \mbox{O$^+$:\,CO$^{+}$} break-up channel of CO$_2^{2+}$ in collisions with Ar$^{4+}$ and Ar$^{6+}$ projectiles, as discussed above, and hence their KERDs are not shown. This observation is consistent with that of Tezuka \emph{et al.} \cite{Tezuka_2018}.  

\begin{figure*}
\centering
\includegraphics[width=0.8\textwidth]{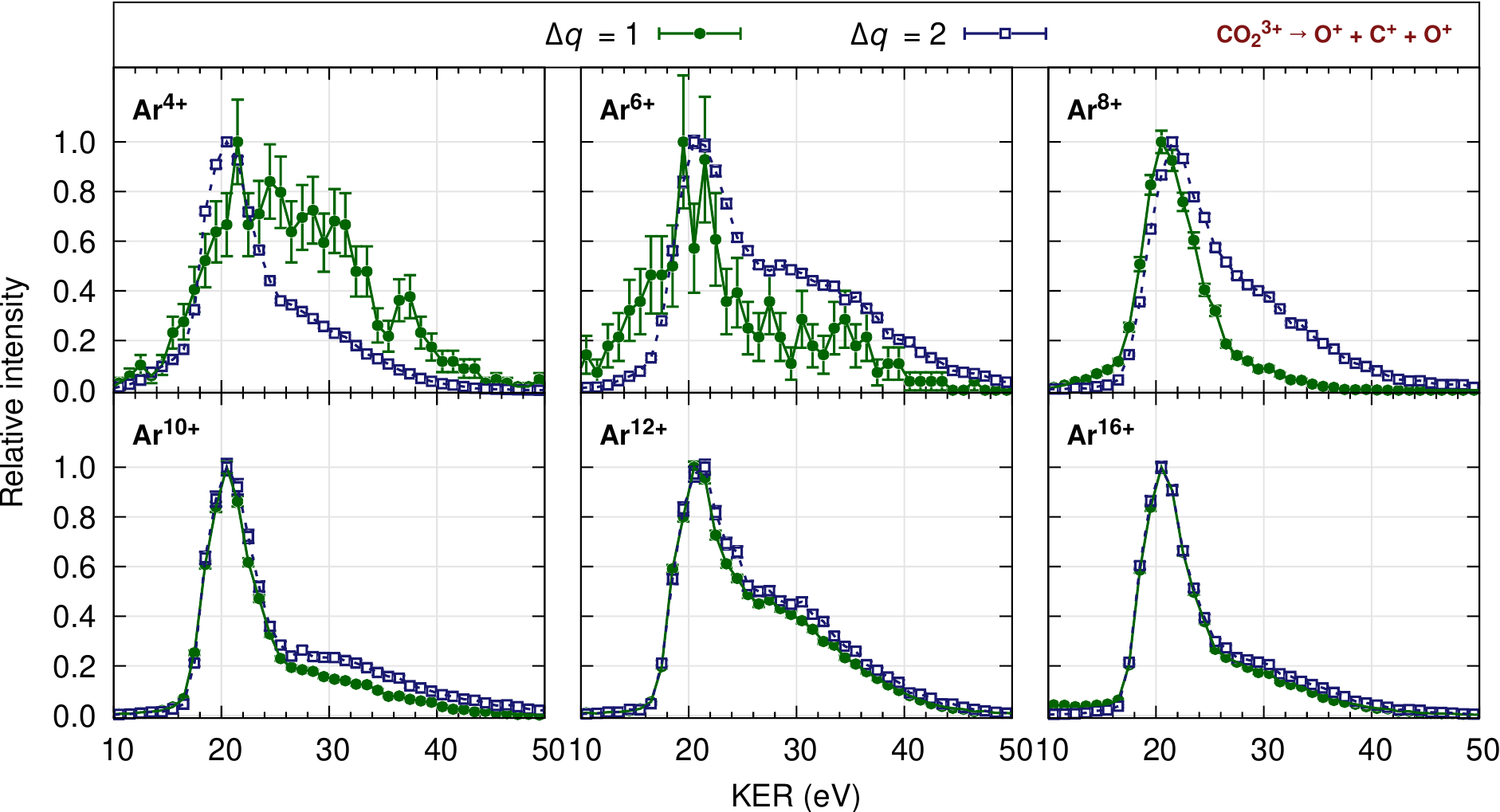}
\caption{Same as Fig.\ref{fig:ker22}, but for the \mbox{O$^{+}$:\,C$^{+}$:\,O$^{+}$} break-up channel of CO$_2^{3+}$.}
\label{fig:ker33}
\end{figure*}

We now turn to the three-body ionic break-up channel, \mbox{O$^{+}$:\,C$^{+}$:\,O$^{+}$}, the KERDs for which are shown in Fig.~\ref{fig:ker33}. All KERDs exhibit a pronounced peak near 20~eV, accompanied by a broad shoulder above 25~eV and an extended high-energy tail. A detailed discussion on the origin of these features can be found in Ref.~\cite{Akash_2026} and references therein. In the present work, we restrict our attention to the influence of capture-associated processes on the KERDs.

Two deviations are seen in the KERDs for Ar$^{4+}$ and Ar$^{6+}$ impact in comparison to the rest of the projectiles.  While in all other projectiles the intensity in the high KER region (beyond $\sim 22$~eV) of the KERD is larger for \mbox{$\Delta q =2$} compared to \mbox{$\Delta q =1$}, it is exactly opposite in the case of Ar$^{4+}$.  The higher intensity for large KER for \mbox{$\Delta q = 2$} relative to that for \mbox{$\Delta q = 1$} is seen most clearly in Ar$^{6+}$ and Ar$^{8+}$ impact and the difference diminishes with increasing projectile charge.  A similar enhancement of the high-KER component for \mbox{$\Delta q=1$} relative to \mbox{$\Delta q=2$} has been reported for two-body fragmentation of N$_2$ \cite{Ehrich_2003} and three-body fragmentation of CO$_2$ \cite{Kumar_2024_a} with low charged projectiles (\mbox{$q \le 3$}). The other deviation is the enhancement of the low KER component (around 15~eV) for Ar$^{6+}$ impact (also weakly for Ar$^{8+}$ impact) for \mbox{$\Delta q=1$} relative to \mbox{$\Delta q=2$} which is in contrast to what is observed for other projectiles.

\subsection{Ionic Fragmentation of CO$_2^{4+}$}
\label{sec:frag4}

For CO$_2^{4+}$, we have observed only triple ionic fragmentation channels, \mbox{O$^{+}$:\,C$^{2+}$:\,O$^{+}$} and \mbox{O$^{2+}$:\,C$^{+}$:\,O$^{+}$}. These break-up channels have recently been investigated by our group \cite{Sumit_2021}. In the present work, these channels are observed for both \mbox{$\Delta q= 1$} and \mbox{$\Delta q= 2$}, but only for \mbox{$q\ge 10$}. For both channels, the KERDs  corresponding to \mbox{$\Delta q= 1$} and \mbox{$\Delta q= 2$} are, by and large, similar for \mbox{$q \ge 10$} and hence are not shown.
 
\section{INTERPRETATION USING REACTION WINDOWS}
\label{sec:interpret}

\begin{figure*}
	\centering
	\includegraphics[width=0.8\textwidth]{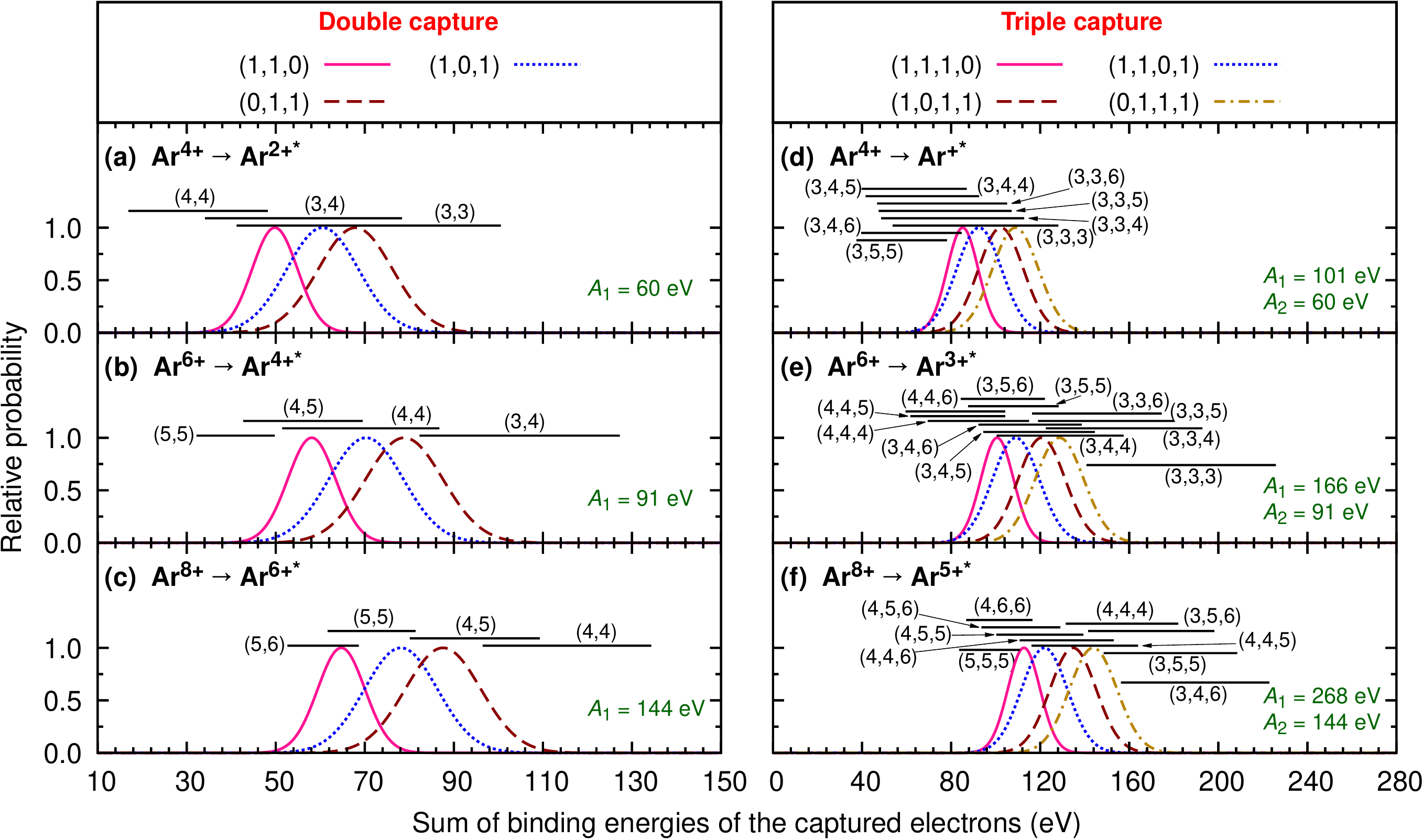}
	\caption{Reaction windows calculated using the ECOBM for different capture strings for \mbox{Ar$^{q+}$--\,CO$_2$} collisions involving double and triple electron capture.  The probability distribution for each window is shown with the peak value arbitrarily set at unity. Horizontal line segments indicate the calculated binding energy ranges corresponding to various \mbox{($n_1l',n_2l''$)} and \mbox{($n_1l',n_2l'',n_3l'''$)} electronic configurations of the scattered projectile and labeled by their respective principal quantum numbers.  Single and double autoionization thresholds for the scattered projectile are given as $A_1$ and $A_2$, respectively.  For \mbox{$q=10,12,16$}, the thresholds are to the far right of the windows, and therefore of no significance.}
	\label{fig:rw_2_3_cap}
\end{figure*}

\subsection{Broad Features of the KERDs}
\label{sec:interpret_broad}

We interpret the KERDs using reaction windows calculated within the framework of the ECOBM, for each degree of target ionization and the number of captured electrons. The difference in target excitation due to \mbox{$r$C$(r-1)$A$_\text{P}$} and \mbox{$r$C$(r-2)$A$_\text{P}$} processes can be inferred by comparing the ranges of the reaction windows relative to the $s$-fold projectile autoionization thresholds, $A_s$. A process is energetically allowed if the sum of the binding energies of the captured electrons, $E^{(j)}$ (See Sec.~\ref{sec:reacwin}), is lower than $A_s$. The presence of at least one threshold within the reaction windows indicates a difference in the degree of target excitation under the two processes. If the ranges of $E^{(j)}$ are always lower than all the relevant $A_s$ for all $j$, all windows can contribute to both, \mbox{$r$C$(r-1)$A$_\text{P}$} and \mbox{$r$C$(r-2)$A$_\text{P}$}, processes. As a result, distinctions in the associated target excitation arising from these processes are expected to diminish. In such cases, the KERDs corresponding to \mbox{$\Delta q=1,2$} are expected to be nearly identical. This situation arises within the model for \mbox{$q\ge 8$} for \mbox{$r=2$} and \mbox{$q \ge 10$} for \mbox{$r = 3, 4$}.

As seen in Fig.~\ref{fig:ker22} and Fig.~\ref{fig:ker33}, the KERDs for CO$_2^{2+}$ for \mbox{$q \ge 8$} and the KERDs for CO$_2^{3+}$ for \mbox{$q\ge 10$} are nearly the same for \mbox{$\Delta q=1,2$}, while for smaller values of $q$ they differ. As noted earlier, CO$_2^{4+}$ fragmentation is observed largely only for \mbox{$q \ge 10$} and the corresponding KERDs for \mbox{$\Delta q=1,2$} are nearly identical.

We therefore focus on KERDs of CO$_2^{2+}$ and CO$_2^{3+}$ for \mbox{$q \le 8$}.  For these cases reaction windows are shown in Fig.~\ref{fig:rw_2_3_cap}. Also shown are the calculated range of binding energies corresponding to various \mbox{($n_1l',n_2l''$)} and \mbox{($n_1l',n_2l'',n_3l'''$)} electronic configurations of the scattered projectile.

\subsubsection{CO$_2^{2+}$}

The formation of CO$_2^{2+}$ is predicted to be via the 2C1A$_\text{P}$ process for all projectile charge states considered here, consistent with the experimentally observed \mbox{$\Delta q=1$} channel. For \mbox{$q=4$}, nearly half of the reaction window corresponding to the (1,0,1) string and the majority of that corresponding to the (0,1,1) string lie below $A_1$ (see Fig.~\ref{fig:rw_2_3_cap}(a)). For \mbox{$q=6$}, a substantial portion of the (0,1,1) string lies below $A_1$ (see Fig.~\ref{fig:rw_2_3_cap}(b)). Under these conditions, ECOBM predicts higher target excitation via a 2C process compared to the 2C1A$_\text{P}$ process for both these projectiles. This is consistent with the observed enhancement of the signal at high KER values for \mbox{$\Delta q=2$} relative to \mbox{$\Delta q=1$}.

For Ar$^{4+}$ impact  the populated configurations of the scattered Ar$^{2+*}$ ion are \mbox{($3l',3l''$)}, \mbox{($3l',4l''$)}, and \mbox{($4l',4l''$)}. While all of these configurations contribute to both 2C1A$_\text{P}$ and 2C processes, the latter is expected to have a relatively larger contribution from low orbital angular momentum ($l$) subshells of the \mbox{($3l',3l''$)} and \mbox{($3l',4l''$)} configurations. Similarly for Ar$^{6+}$, the 2C process is expected to have an enhanced contribution from low-$l$ subshells of the \mbox{($3l',4l''$)} configurations of the scattered Ar$^{4+*}$ ion.  For \mbox{$q \ge 8$}, all populated electronic configurations of the scattered projectile are expected to contribute to both 2C1A$_\text{P}$ and 2C processes. As $q$ increases from 4 to 16, the principal quantum number $n$ of the populated states increases systematically, indicating capture into increasingly higher excited states of the scattered Ar$^{(q-2)+*}$ projectile.

\subsubsection{CO$_2^{3+}$}

The formation of CO$_2^{3+}$ with \mbox{$\Delta q=2$} is predicted to be predominantly via the 3C1A$_\text{P}$ process for all projectiles in the present work. On the other hand, the formation of CO$_2^{3+}$ with \mbox{$\Delta q=1$} for \mbox{$q \ge 8$} is predicted to be predominantly via the 3C2A$_\text{P}$ process. For \mbox{$q=4$}, double autoionization of the scattered Ar$^{+*}$ ion is negligible for all capture strings. Consequently, for Ar$^{4+}$, the experimentally observed \mbox{$\Delta q=1$} channel may arise either from 1C2I$_\text{T}$ or from single-capture followed by double autoionization of the residual target ion (1C2A$_\text{T}$). Similarly, for \mbox{$q=6$}, double autoionization of the scattered Ar$^{3+*}$ ion is allowed only over a limited region of the (1,1,1,0) and (1,1,0,1) capture strings. Therefore, for Ar$^{6+}$, the experimentally observed \mbox{$\Delta q=1$} channel may contain significant contributions from 1C2I$_\text{T}$ and 1C2A$_\text{T}$, in addition to the 3C2A$_\text{P}$ process. For these projectiles, additional processes (apart from $r$C$s$A$_\text{P}$) must then be invoked to explain the observations. The interpretation of the KERDs for Ar$^{4+}$ impact and Ar$^{6+}$ impact is presented in the following subsection.

For \mbox{$q=8$}, while the reaction windows for all capture strings lie above $A_1$, nearly half of the reaction window corresponding to the (0,1,1,1) string and the majority of that corresponding to the (1,0,1,1) string lie below $A_2$ (see Fig.~\ref{fig:rw_2_3_cap}(f)). Under these conditions, ECOBM predicts that the 3C1A$_\text{P}$ process results in a higher degree of excitation to the target compared to the 3C2A$_\text{P}$ process. Following the same reasoning as for the two-body fragmentation of CO$_2^{2+}$, this excess excitation can be correlated with the weak enhancement of the high-KER component observed for \mbox{$\Delta q=2$} relative to \mbox{$\Delta q=1$} in the KERDs corresponding to the \mbox{O$^+$:\,CO$^{2+}$} break-up channel (as mentioned in Section~\ref{sec:frag3}). In addition, rovibronic excitation of the CO$^{2+}$ molecular fragment may store a fraction of the available energy, thereby reducing the observable KER and leading to only a modest enhancement in the high-energy tail for \mbox{$\Delta q=2$}. However, this rovibronic energy is released upon complete atomization of CO$_2^{3+}$ into \mbox{O$^{+}$:\,C$^{+}$:\,O$^{+}$}, which correlates with the considerable enhancement of the high-KER component observed for \mbox{$\Delta q=2$} relative to \mbox{$\Delta q=1$} in the corresponding KERDs (see Fig.~\ref{fig:ker33}).

For \mbox{$q \ge 10$}, all electronic configurations contribute to both 3C1A$_\text{P}$ and 3C2A$_\text{P}$ processes. In contrast, for the Ar$^{8+}$ impact, the configurations \mbox{($3l',4l'',6l'''$)}, \mbox{($3l',5l'',5l'''$)}, and \mbox{($3l',5l'',6l'''$)} contribute exclusively to the 3C1A$_\text{P}$ process. In addition, 3C1A$_\text{P}$ receives enhanced contributions from the low-$l$ subshells of the \mbox{($4l',4l'',4l'''$)}, \mbox{($4l',4l'',5l'''$)} and \mbox{($4l',4l'',6l'''$)} configurations of the scattered Ar$^{5+*}$ ion.

\subsection{The $r$C$s$A$_\text{T}$ process in CO$_2^{3+}$ formation}
\label{sec:interpret_CA_T}

As noted above, two features stand out in the KERDs for \mbox{O$^+$:\,C$^{+}$:\,O$^{+}$} fragmentation channel of CO$_2^{3+}$ in the case of Ar$^{4+}$ and Ar$^{6+}$ impact.  These are the enhancement of high KER ($> 22$~eV) signal for Ar$^{4+}$ impact and the enhancement of the low KER component (around 15~eV) for Ar$^{6+}$ impact. In both instances the KER signal for \mbox{$\Delta q=1$} is higher than for \mbox{$\Delta q=2$}, opposite to that seen for other projectiles. 

\begin{figure}
	\centering
	\includegraphics[width=0.9\columnwidth]{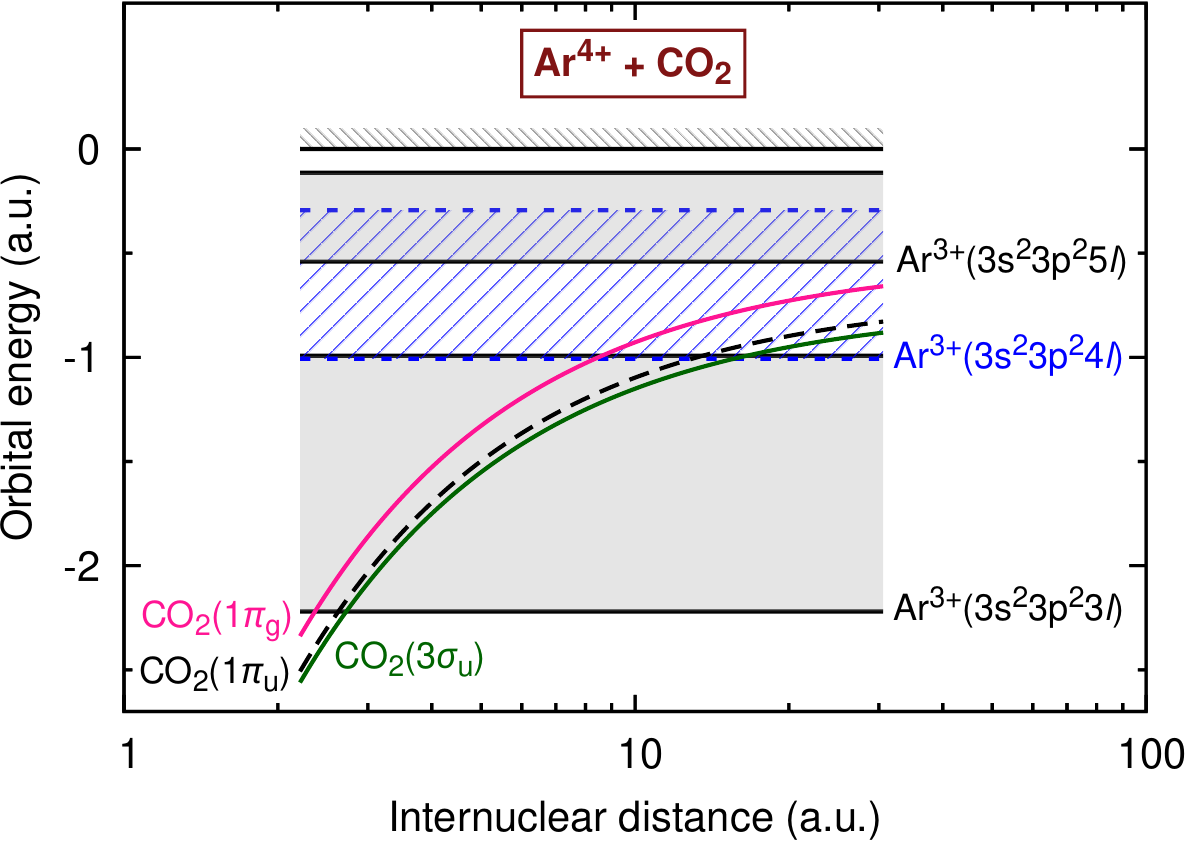}
	\caption{Orbital energies for the Ar$^{4+}$--\,CO$_2$ collision system (see text for details).}
	\label{fig:orben}
\end{figure}

For Ar$^{4+}$--\,CO$_2$ collisions, ECOBM calculations predict onset of double autoionization of CO$_2^{+}$ following single-capture (1C2A$_\text{T}$ process) when the first four electrons, starting from the outermost, are recaptured into excited states of the target while the fifth electron is captured by the projectile, corresponding to the string (0,0,0,0,1). The ground-state electronic configuration of neutral CO$_2$ is \mbox{$(1\sigma_\text{g})^2$$(1\sigma_\text{u})^2$$(2\sigma_\text{g})^2$$(3\sigma_\text{g})^2$$(2\sigma_\text{u})^2$$(4\sigma_\text{g})^2$$(3\sigma_\text{u})^2$$(1\pi_\text{u})^4$$(1\pi_\text{g})^4$}. The string (0,0,0,0,1) corresponds to single-capture from the $1\pi_\text{u}$ orbital of CO$_2$. To assess whether capture from inner-valence orbitals is plausible, we analyze the orbital energy diagrams for the \mbox{Ar$^{4+}$--\,CO$_2$} collision system in Fig.~\ref{fig:orben}, as was done for the investigation of He$^{2+}$--CO collisions \cite{Folkerts_1997,Sobocinski_2001}.

In Fig.~\ref{fig:orben}, energies of electrons in the three outermost orbitals of CO$_2$ are shown for the incident channel and the orbital energies of the scattered Ar$^{3+}$ ion are shown in the exit channel. Assuming resonant capture, electrons are transferred at crossings between the Stark-shifted target orbitals and projectile orbitals. Thus, single-capture from the inner-valence orbitals $1\pi_\text{u}$ and $3\sigma_\text{u}$ is possible at small internuclear distances, where strong radial couplings exist. At such small distances, the perturbation exerted on the target is substantial and can lead to the formation of CO$_2^{+}$ in excited states above the double autoionization threshold, resulting in triple ionization via a 1C2A$_\text{T}$ process with \mbox{$\Delta q =1$}. Compared to the 3C1A$_\text{P}$ process (which will result in triple ionization with \mbox{$\Delta q =2$}), which predominantly involves capture from the outermost $1\pi_\text{g}$ orbital, the 1C2A$_\text{T}$ process involves smaller impact parameters and stronger target perturbation. This explains the enhancement beyond 22~eV KER observed for the \mbox{$\Delta q=1$} channel (see Fig.~\ref{fig:ker33}). This enhancement may also occur in the case of 1C2I$_\text{T}$ process \cite{Kumar_2024_a}, which may be verified by Auger electron spectroscopy of the target combined with projectile-charge-resolved measurements. 

The 15~eV KER feature observed for Ar$^{6+}$ impact (for \mbox{$\Delta q=1$}) has been reported earlier.  In regimes where \mbox{$r$C$i$I$_\text{T}$} or DI dominate, the feature was attributed primarily to sequential breakup of CO$_2^{3+}$ \cite{Wang_2015,Sumit_2022}. Where true capture or $r$C$s$A$_\text{P}$ processes dominate, this feature was attributed mainly to concerted breakup, originating from fragmentation of the low-lying $^2\Pi_\text{g}$ and $^{2,4}\Pi_\text{u}$ states correlating with the second \mbox{(D$_2$: C$^+$($^2$P) + O$^+$($^4$S) + O$^+$($^2$D))} and third dissociation limits \mbox{(D$_3$: C$^+$($^4$P) + 2 O$^+$($^4$S)} \cite{Akash_2026}. Importantly, both the D$_2$ and D$_3$ limits correlate with the $^2\Pi_\text{g}$ and $^{2,4}\Pi_\text{u}$ states in non-linear molecular configurations \cite{Akash_2026}.

If these states are accessed via the 1C2I$_\text{T}$ or 2C1I$_\text{T}$ mechanisms, a vertical Franck–Condon transition from the linear geometry of CO$_2$ can be reasonably assumed given the short collision time of $\approx$ 1~fs. Subsequent evolution of the $^2\Pi_\text{g}$ and $^{2,4}\Pi_\text{u}$ states from this vertical transition point is energetically favored toward the first dissociation limit \mbox{(D$_1$: C$^+$($^2$P) + 2 O$^+$($^4$S))}, which does not produce the 15~eV feature.

If, on the other hand, the 1C2A$_\text{T}$ contributes significantly to \mbox{$\Delta q=1$}, the initial single capture leaves the CO$_2^{+**}$ molecular ion in excited states, subsequent double autoionization of which leads to CO$_2^{3+}$.  Since autoionization lifetimes typically range from a few femtoseconds to a few tens of femtoseconds, the nuclear wave packet can evolve on the potential energy surface of CO$_2^{+**}$. This allows the system to explore non-linear configurational phase space prior to autoionization. The subsequent population of the $^2\Pi_\text{g}$ and $^{2,4}\Pi_\text{u}$ states in non-linear configurations can then correlate with the D$_2$ and D$_3$ dissociation limits, giving rise to the observed 15~eV feature.

For Ar$^{6+}$ projectiles, the absence of the 15~eV feature in the KERDs associated with the \mbox{$\Delta q=2$} channel implies that 2C1A$_\text{T}$ is insignificant relative to the 3C1A$_\text{P}$ process. This is consistent with the reaction windows (see Fig.~\ref{fig:rw_2_3_cap}\textcolor{blue}{(e)}), which indicate that 3C1A$_\text{P}$ is strongly allowed. Additionally, if the \mbox{$\Delta q=1$} channel were dominated by 1C2A$_\text{T}$ and 1C2I$_\text{T}$ processes, one would expect variations in KERDs for \mbox{$\Delta q = 1,2$} to be similar to those observed for Ar$^{4+}$ projectiles. The observed trend is opposite, leading to the inference that the \mbox{$\Delta q=1$} channel is dominantly fed by the 3C2A$_\text{P}$ process. The absence of the 15~eV feature in the KERDs for \mbox{$q \ge 8$} and \mbox{$\Delta q = 1,2$} strengthens the assumption that $r$C$s$A$_\text{P}$ processes are the predominant ionization mechanisms for $q \ge 8$.

\section{CONCLUSIONS}
\label{sec:concl}

Capture-associated processes governing the fragmentation dynamics of CO$_2^{n+}$, \mbox{($2 \le n \le 4$)}, have been analyzed, with particular emphasis on the roles of projectile charge $q$ and charge change \mbox{$\Delta q$}. Reaction windows calculated within the ECOBM framework provide a qualitative basis for interpreting the observed trends in the KERDs. In most cases, multifold autoionization of the scattered projectile following multielectron capture is found to be responsible for the various post-collision charge states of the projectile. 

For very highly charged projectiles, where the reaction windows lie far above the corresponding thresholds, the measured KERDs for \mbox{$\Delta q=1,2$} are found to be nearly identical across all investigated ionic fragmentation channels of CO$_2^{n+}$ (\text{$2 \le n \le 4$}). Since in this case, the full extent of the reaction windows contribute to all the relevant processes, distinctions in the associated target excitation arsing from these processes are expected to diminish, in agreement with observed KERDs.

Where the reaction windows permit both \mbox{$\Delta q=1$} and \mbox{$\Delta q=2$}, but with a relatively larger available energy range for the latter, the predictions of ECOBM are consistent with experimental observations. In such cases, the KERDs for \mbox{$\Delta q=2$} exhibit an enhancement in the high KER region relative to the KERDs for \mbox{$\Delta q=1$}.

Where the reaction windows allow only \mbox{$\Delta q=2$} (or allow \mbox{$\Delta q=1$} only weakly), the trends in the KERDs for \mbox{$\Delta q=1,2$} differ from those observed in the preceding scenarios. Processes other than capture followed by projectile autoionization must be invoked to account for the fragmentation corresponding to \mbox{$\Delta q=1$}. Together they explain the observed differences in the KERDs for \mbox{$\Delta q=1,2$}.

For intermediate values of projectile charge, a high degree of target excitation (inferred in terms of large signal for high KER) is found to be associated with capture to excited states of the scattered projectile (inferred from the number of projectile autoionizations). For very highly charged projectiles, the degree of target excitation appears to become relatively insensitive to projectile autoionization, when the number of autoionized electrons is small.


\acknowledgments

The authors thank the Dept.\ of Science and Technology, Science and Engineering Research Board (India) for generous funding via grant No. 30116294, which enabled the setting up of the EBIS/A ion source. They would also like to acknowledge technical assistance from Mr.\ Nilesh Dumbre of IISER Pune in maintenance of the ion source. A. S. and V.P. acknowledge the award of fellowship by CSIR.

\bibliography{Capture_Role_11}

\end{document}